\documentclass[fleqn,usenatbib]{mnras}

\usepackage[T1]{fontenc}

\DeclareRobustCommand{\VAN}[3]{#2}
\let\VANthebibliography\thebibliography
\def\thebibliography{\DeclareRobustCommand{\VAN}[3]{##3}\VANthebibliography}

\usepackage{graphicx}	
\usepackage{amsmath}	

\renewcommand\ion[2]{#1$\;${\small\rmfamily{#2}}\relax}

\title[An Empirical Calibration of the Helium Abundance]{An Empirical Calibration of the Helium Abundance in HII Regions based in Literature and CALIFA Survey data}

\author[Valerdi, M. et al.]{%
M. Valerdi$^{1}$\thanks{E-mail: mvalerdi@astro.unam.mx}, 
J. K. Barrera-Ballesteros,$^{1}$
S. F. S\'anchez,$^{1}$
C. Espinosa-Ponce,$^{1}$ \newauthor
L. Carigi$^{1}$ 
and  A. Mej\'ia-Narv\'aez,$^{1}$
\\
$^{1}$Instituto de Astronom\'ia, Universidad Nacional Aut\'onoma de M\'exico, AP 70264, 04510 Mexico City, Mexico}

\date{Accepted XXX. Received YYY; in original form ZZZ}

\pubyear{2020}

\usepackage{newtxtext,newtxmath}

\begin{document}
\label{firstpage}
\pagerange{\pageref{firstpage}--\pageref{lastpage}}
\maketitle

\begin{abstract}
Helium is the second most common chemical species in the Universe. The study of helium abundance has the potential to unravel the chemical evolution of and within galaxies. In this study, we provide an empirical calibration for the singly ionized helium abundance: $12+\log_{10}({\rm He}^+/{\rm H}^+)$, based on the emission line flux ratio He$_{\lambda5876}$/H$\alpha$ from Galactic and extragalactic \ion{H}{II} regions compiled from the literature. Based on this calibrator, we explore for the first time the helium abundance in a large sample of \ion{H}{II} regions located in galaxies representative of the nearby Universe from the CALIFA survey. Furthermore, this calibrator allows us to explore the variations of the helium abundance with respect to the oxygen abundance. The observed trends are in agreement with a change in the chemical enrichment with mass/oxygen abundance similar to the one observed due to the inside-out model in a MW-galaxy (highlighting the connection between resolved and global trends in galaxies). Our calibrator provides an empirical proxy to estimate the helium abundance at kpc scales as well as to constrain chemical evolutionary models.
\end{abstract}



\section{Introduction}

The formation of the elements is a key to understand the evolution of the Universe. In particular, the formation of helium has been fundamental to the study of cosmology and the chemical evolution of galaxies. The majority of the cosmic helium was produced just after the Big Bang, during the primordial nucleosynthesis phase \citep{Kobayashi2020}. Afterward, the first stars that formed contained only hydrogen and helium. The heavier elements have been formed from nuclear reactions in the interior of successive generations of stars \citep[e.g][]{Bromm2004,Schneider2006}. 

In the present helium constitutes about $24-25\%$ of the baryonic mass in the Universe and it is the second most abundant element. Helium is produced by hydrostatic nucleosynthesis in the interior of stars of all initial masses. Massive stars produce this element through the CNO cycle and low-mass stars via the proton-proton chain \citep[e.g.][]{Weaver1980,Arnett1985}. Finally, there is an amount of helium that enriches the ISM through stars that ejected it. This production depends on the progenitor star's initial mass, metallicity, stellar rotation, and stellar winds \citep[e.g][]{Peimbert1980,Peimbert1986}.

In principle, due to the helium nature, one might expect that their abundance to be well determined. However, it is not possible to observe all ionization stages, particularly when it is in the neutral form. Due to the high efficiency of the absorption of ultraviolet radiation by hydrogen, the boundary between ionized and neutral gas is very well defined. For this reason, helium may exist in its neutral form in the outermost parts of these regions. The neutral helium cannot be measured, for this reason, the total helium abundance can only be estimated in \ion{H}{II} regions where all helium is ionized \citep[high excitation \ion{H}{II} regions, e.g. Orion nebula][]{Esteban2004}. To obtain the total abundance of an element one would have to sum over all their ionic stages (and correct for the neutral component). However, for most elements, we cannot observe all the ionic stages, and the helium is not the exception. To solve this problem it is usually used Ionization correction factors (ICFs) \citep[e.g.][]{Peimbert1977,Stasinska1978,Vilchez1989,PerezM2007,Delgado2014,Amayo2020}.

Observationally, the spectra of \ion{H}{II} regions only show recombination lines of \ion{He}{I} and sometimes of \ion{He}{II}. However, as we mentioned before, it is possible that it is found in a neutral form in the external areas of these regions. \citet{Vilchez1989,Deharveng2000-II} suggested that within the \ion{H}{II} regions with low and medium ionization degrees the presence of neutral helium is important and the ${\rm ICF(He)}>1$. However, for objects with a high ionization degree, we can consider the presence of neutral helium within the nebula to be negligible. To obtain the total helium abundance, there are several ICF schemes for helium available in the literature and based on different ionic ratios \citep[e.g.,][]{1977MNRAS.179..217P,1992RMxAA..24..155P,1983ApJ...273...81K,2003A&A...404..545Z,Delgado-Inglada2014}.

To obtain the ionic abundance (He$^+$/H$^+$), it is necessary to consider some parameters such as (i) recombination coefficients for helium and hydrogen lines; (ii) correction due to effects of collisional excitation; (iii) correction due to effects of optical depth. In addition, there are several sources of uncertainty in the determination of the helium abundance. The most important one is related to the ionization structure of the nebulae (respect to the ICF). The systematic effects also introduce uncertainties as well. These can be divided in three broad categories: i) uncertainties in the atomic data for \ion{He}{I} (recombination emissivities, collisional enhancement coefficients), ii) observational effects (interstellar extinction correction, underlying stellar absorption), and iii) possible deviations from the adopted ionization-bounded homogeneous \ion{H}{II} region standard model \citep[deviations from case B radiative recombination theory, radiative transfer in HeI lines, temperature fluctuations, etc.; e.g.,][]{Izotov1997}.

The main goal of this study is to obtain a calibrator to estimate the single ionized helium abundance using the best compilation of \ion{H}{II} regions available in the literature, and compare with helium abundances obtained from the catalog of \ion{H}{II} regions of CALIFA. The helium abundance is usually measured using various emission lines in the optical range and using spectra with high spectral resolution, thus having reliable observations on weak lines. The helium abundance has been studied in several works \citep[e.g.][]{Izotov2007,Peimbert2007,Aver2015,Eduardo2020}, the largest sample is reported by \citet{Izotov2007}, with $\sim350$ \ion{H}{II} regions. Most of these works aim to study the primordial helium abundance and therefore seek a highly accurate determination. In this study, we derive an empirical calibration for the singly ionized helium abundance, and we derive it in one of the largest sample of \ion{H}{II} (5386 \ion{H}{II} regions). This calibrator is important because helium has not frequently used to trace chemical evolution. However, this study could help to constrain the chemical evolution models, and consequently, we could understand better the chemical enrichment process in galaxies. In addition our sample was extracted from an statistical significant sample of galaxies representative of the population in the nearby Universe.

The content of the article is distributed as follows: In Section 2, we describe two different datasets on which our study is based: I) the compilation of Galactic and extragalactic \ion{H}{II} regions available in the Literature with helium abundance determinations, II) the largest catalog of \ion{H}{II} regions with spectroscopic information extracted from the CALIFA dataset \citep{Espinosa-Ponce2020}. In Section 3, we present the details on the analysis procedures and the determination of helium abundance. Finally, Section 4, summarizes the discussion and conclusions of this work.

\section{Sample}

The current study is based on two different datasets. The first one is a compilation of all the available \ion{H}{II} regions with helium abundance in the literature. The second one comprises the largest catalog of \ion{H}{II} regions with spectroscopic information extracted from the CALIFA dataset \citep{Espinosa-Ponce2020}. In Sec \ref{sec:literature} we descibe the compilation from the literature while in Sec \ref{sec:califa_HII} we describe the CALIFA survey, and in Sec \ref{sec:califa_HII} we describe the \ion{H}{II} regions catalog derived from the CALIFA survey.

\subsection{Compilation of helium abundances in \ion{H}{II} regions}
\label{sec:literature}

We perform a search in the literature of \ion{H}{II} regions in order to compile a sample of them with a good estimation of helium abundances. We search those targets in previous works with estimations of helium abundances using the \ion{He}{I} lines in the optical range that include $\lambda3819$, $\lambda4026$, $\lambda4471$, $\lambda4713$, $\lambda4922$, $\lambda5876$, and $\lambda6677$. The final compilation comprises 174 \ion{H}{II} regions from 8 different works in the literature. Table \ref{tab:literature} shows the references from these studies. We summarize here the main properties of those samples.

\citet{Izotov1997} observed high-quality spectra of 45 supergiants \ion{H}{II} regions in low-metallicity blue compact galaxies (BCGs) with the aim to compute the primordial helium abundance. The observations were made with the Ritchey-Chrétien RC2 spectrograph at the Kitt Peak 4m Mayall Telescope. The spectral range is $3500-7500$\AA\, and a spectral resolution of $\sim5$\AA. They computed ionized helium abundance in each \ion{H}{II} region through the helium emission-lines: $\lambda4471$, $\lambda5876$, and $\lambda6678$, as obtained by a self-consistent procedure. They used two different sets of theoretical recombination \ion{He}{I} line emissivities: one by \citet{Brocklehurst1971} and another one by \citet{Smits1996}. To obtain the total helium abundance, they estimated the radiation softness parameter ($\eta$), to correct for neutral helium \citep{Vilchez1988}. 

\citet{Deharveng2000-II} analyzed spectra of 34 Galactic \ion{H}{II} regions. The observations were made with ESOP, Fabry-Perot spectrophotometer at the 1.5m telescope of the Observatorio Astronómico Nacional at San Pedro Mártir. The spectral range is $3400-7000$\AA\, with a spectral resolution of $\sim1$\AA. To calculate ionized the helium abundance they used the \ion{He}{I} $\lambda5876/{\rm H}\beta$ ratio, corrected for extinction and they assumed a two temperature \ion{H}{II} region model. They used the H$\beta$ emissivity from \citet{Storey1995} and the \ion{He}{I} $\lambda5876$ emissivity from \citet{Benjamin1999} (which include the effect of collisional excitation from the 2S levels, and thus eliminate the need to correct for collisional enhancement of the \ion{He}{I} lines). Through the O$^{++}$/O ratio, they observed that most of the sample has values ${\rm O}^{++}/{\rm O}<0.4$, and therefore contain neutral helium. When the ratio value is $0.70<{\rm O}^{++}/{\rm O}<0.85$, they assumed that ${\rm He}^+/{\rm H}^+={\rm He}/{\rm H}$. Finally, they concluded that an O6.5 star may ionize an \ion{H}{II} region that contains a significant fraction of neutral helium.

\citet{Peimbert2003,Peimbert2005,Peimbert2012,Valerdi2019,Valerdi2021} studied small samples (from 1 to 5) of \ion{H}{II} regions in comparison with the previous ones. Their main goal was to estimate the primordial helium abundance. The spectra analyzed in these works have been observed with the Ultraviolet Visual Echelle Spectrograph (UVES) at the 8m VLT Kueyen Telescope. The spectral range was $3100-10360$\AA\, and the FWHM resolution was $\Delta\lambda\sim\lambda/8800$. Data were observed at at the VLT Telescope using Spectrograph FORS1. Three grism settings were used: GRIS-600B+12, GRIS-600R+14 with filter GG435, and GRIS-300V in addition to the filter GG375, the spectral range is $3450-5900$\AA\ with a spectral resolution of $\lambda/\Delta\lambda\sim1300$, $5350-7450$\AA\ with a spectral resolution of $\lambda/\Delta\lambda\sim1700$, and $3850-8800$\AA\ with a spectral resolution of $\lambda/\Delta\lambda\sim700$, respectively. To derive the He$^+$/H$^+$ value of a region, they used the following \ion{He}{I} lines in the optical range: $\lambda3889$, $\lambda4026$, $\lambda4387$, $\lambda4471$, $\lambda4713$, $\lambda4922$, $\lambda5876$, $\lambda6677$, and $\lambda7065$. In order to optimize the data, they used the \texttt{Helio14} code \citep{Peimbert2012}, which is an extension of the maximum likelihood method to search for the physical and chemical conditions ($T_0$, $t^2$, $\tau_{3889}$, and He$^+$/H$^+$) that would give them the best simultaneous fit to all the measured lines. They used recombination coefficients were those by \citet{Storey1995} for H and those by \citet{Smits1996} and \citet{Benjamin1999} for He. Using the results from \citet{Sawey1993} and \citet{Kingdon1995}, they estimated the collisional contribution, and using the computations by \citet{Benjamin2002} they estimated the optical depth effects. To estimate the helium total abundance, they used different ICF(He$^0$) expressions to correct for neutral helium.

\citet{Eduardo2020} determined the radial abundance gradient of helium in the Milky Way disc from spectra of 19 \ion{H}{II} regions and ring nebulae surrounding massive O-type stars. The data were compiled from previous works of the same group. The spectra were obtained with different telescopes and instruments: (i) The UVES  spectrograph at the VLT; the spectral range was $3100-10400$\AA\, and the resolution at a given wavelength is given by $\Delta\lambda\sim\lambda/8800$, that corresponds to an average FWHM$\sim0.265$\AA; (ii) The Optical System for Imaging and low-Intermediate-Resolution Integrated Spectroscopy (OSIRIS) spectrograph at the 10.4m Gran Telescopio Canarias (GTC); two grism were used: R1000B and R2500V, the spectral range is $3640-7870$\AA\ with a spectral resolution of 6.52\AA,  and $4430-6070$\AA\ with a spectral resolution of 2.46\AA, respectively; and finally (iii) the Magellan Echellette (MagE) spectrograph at the 6.5m Clay Telescope; the spectral range is $3100-10 000$\AA\, and a spectral resolution of $\lambda/\Delta\lambda\sim4000$. They determined the He$^+$/H$^+$ abundance with the code \texttt{PYNEB} \citep{pyneb2015} using all or several of the following list of\ion{He}{I}lines: $\lambda3614$, $\lambda3889$, $\lambda3965$, $\lambda4026$, $\lambda4121$, $\lambda4388$, $\lambda4438$, $\lambda4471$, $\lambda4713$, $\lambda4922$, $\lambda5016$, $\lambda5048$, $\lambda5876$, $\lambda6678$, $\lambda7065$, $\lambda7281$, and $\lambda9464$. The recombination coefficients used were those by \citet{Porter2012,Porter2013} for He lines and \citet{Storey1995} for H lines. To obtain the total helium abundance, they used four different ionization correction factor (ICF; He) schemes. 

Like in the previous explorations \citet{Aver2020} were interested in the study of the primordial helium. They studied 16 low-metallicity extragalactic \ion{H}{II} regions to lower limit the systematic uncertainties in helium abundance determinations. The sample contain 15 objects from \citet{Aver2015} and they added the brightest \ion{H}{II} region in the extremely metal-poor ($\sim3\%$ Z$\odot$) galaxy Leo~P observations. Last one were taken with the optical \& near-IR spectrum, from the UV atmospheric cutoff to $1\mu$m. They used the LBT's Multi-Object Double Spectrograph LBT/MODS, and the LBT Utility Camera in the Infrared LBT/LUCI to obtain a near-IR spectrum, from 0.95 to 1.35 $\mu$m (to measurement \ion{He}{I} $\lambda10830$ emission line).

\begin{table}
\centering
\caption{Bibliographic references to the original works for the compiled helium abundance sample.}
\label{tab:literature}
\begin{tabular}{lc}
\hline
Reference & Number of \ion{H}{II} regions   \\
\hline
\citet{Izotov1997}                              &    58$^{\rm a}$ \\
\citet{Deharveng2000-II}                        &    64$^{\rm b}$ \\
\citet{Peimbert2003,Peimbert2005,Peimbert2012}  &    13$^{\rm a}$ \\         
\citet{Valerdi2019,Valerdi2021}                 &     8$^{\rm a}$ \\         
\cite{Eduardo2020}                              &    15$^{\rm b}$ \\
\cite{Aver2020}                                 &    16$^{\rm a}$ \\
Total                                           &   174           \\
\hline
\multicolumn{2}{l}{$^{\rm a}$ Extragalactic \ion{H}{II} regions} \\ 
\multicolumn{2}{l}{$^{\rm b}$ Galactic \ion{H}{II} regions}  
\end{tabular}
\end{table}

\subsection{Catalog of \ion{H}{II} regions on the CALIFA survey}
\label{sec:califa_HII}

The CALIFA Survey (Calar Alto Legacy Integral Field Area Survey) is an astronomical project to map galaxies with integral field spectroscopy (IFS), to allow detailed studies of the spatially resolved spectroscopic properties of these objects. The data was taken using the PPAK Integral Field Unit (IFU) \citep{Kelz2006}, of the PMAS spectrograph \citep{2005PASP..117..620R} mounted at the Calar Alto 3.5m telescope, with a hexagonal field-of-view of $74$arcsecs$ \times64$arcsecs, and a $100\%$ covering factor by adopting a three pointing dithering scheme. The optical wavelength range is covered from $3700$ to $7000$\AA \citep{Sanchez2012}. CALIFA is the survey that samples galaxies with the largest independent number of spatial elements for the largest field of view, and allows us to study a statistically significant sample of galaxies, representative of the population at the nearby Universe \citep[e.g.][]{Espinosa-Ponce2020,Sanchez2016DR3,Review2020}.

CALIFA provides IFS data for a sample of galaxies in the local Universe ($0.005< z <0.03$). These galaxies were originally selected from the Sloan Digital Sky Survey imaging survey to have a similar projected size, covering any morphological type. The survey has been designed to allow to build two-dimensional maps of the following quantities: I) stellar populations: ages, metallicities and star formation histories; II) ionized gas: two-dimensional distribution of the flux and EW for each emission line, and chemical abundances; and III) kinematic properties: both from stellar and ionized gas component. The analyzed data fulfill the expectations of the original observing proposal. The process of data reduction and analysis is provided by \citep{Sanchez2012,Sanchez2016DR3}.

Along this article we use the catalog of spectral properties of \ion{H}{II} regions obtained by \citet{Espinosa-Ponce2020}. This catalog is based on the integral field spectroscopy (IFS) data of the extended CALIFA sample comprising 941 galaxies. It includes galaxies observed with the same setup extracted either from the original sample and a subset of extended sub-samples comprising galaxies underrepresented in the original sample \citep[e.g., large elliptical galaxies, dwarf galaxies, mergers, etc.][]{Barrera-Ballesteros2015} that fulfill most of the selection criteria (in particular the diameter selected), as described in \citet{Sanchez2016DR3}. The largest of these sub-samples ($\sim150$ galaxies) corresponds to the PISCO survey \citep{Galbany2018}, that include all SN-hosts within the footprint of the CALIFA selection criteria. The \citet{Espinosa-Ponce2020} catalog of spectroscopic properties is the largest catalog of \ion{H}{II} regions. The detection of \ion{H}{II} regions was based on: (i) their shape (clumpy/peaked) structures in the H$\alpha$ emission maps extracted from the datacubes; (ii) the EW of H$\alpha$ ($>6$\AA); (iii) the presence of a fraction of ionizing young stellar populations larger than a 4\% (in the V-band).

To derive these parameters (H$\alpha$ fluxes and EW, and stellar population properties), the CALIFA data were analyzed using the Pipe3D pipeline. This pipeline has been extensively used and described in previous articles \citep{Sanchez2016a,Sanchez2016b}. As a brief summary, it performs an automatic decomposition of the stellar population (using a multi Single Stellar Population modelling), and the emission lines (using both a Gaussian fitting and a momentum analysis), for all the spectra corresponding to each position within the IFS datacubes. For details on the fitting procedure, the currently adopted SSP library, the full set of emission lines analyzed, and the low and high order data products provided we refer the reader to Pipe3D articles \citet{Sanchez2016a,Sanchez2016b}. The final data products of Pipe3D are a set of maps comprising the spatial distribution of the different derived properties. From these maps, \citet{Espinosa-Ponce2020} extracted the \ion{H}{II} region catalog that is the basis of this study. This catalog provides information of 26,408 individual regions including (i) the flux intensities and equivalent widths of 51 emission lines covering the wavelength range between 3745 and 7200\AA{} corrected by dust with the extinction law by \citet{Cardelli1989} and (ii) the corresponding properties of the underlying stellar population. In this catalog, the \ion{H}{II} regions also were corrected by diffuse ionized gas (DIG).

\section{Analysis and results}

\subsection{Helium abundances He$^+$/H$^+$ in the literature \label{sec:He_literature}}

The helium abundance is frequently derived using several emission lines in the optical and NIR wavelength ranges \citep[e.g][]{Izotov1997,Deharveng2000-II,Peimbert2005,Izotov2014,Valerdi2019,Valerdi2021,Aver2020}. As mentioned before, many of these studies were used to determine the primordial helium abundance and therefore was important to compute the helium abundance in all their ionic stages and correct for the neutral component (see Sec. \S~\ref{sec:literature}). 

In particular, the singly ionized helium abundance obtained in those works has been computed using from 2 to 20 \ion{He}{I} emission lines. For each line there is an estimate of He$^+$/H$^+$ abundance and the final value is calculated using the weighted mean. The authors used different software to estimate this abundance, for example, \texttt{PYNEB}\footnote{http://research.iac.es/proyecto/PyNeb}, the \texttt{Helio14} code, and a set of theoretical recombination \ion{He}{I} line emissivities \citep[e.g.][]{Brocklehurst1971,Porter2013}. In addition, to compute helium abundance in these codes, it is necessary to know the electron density, $N_e$ and temperature, $T_e$[\ion{O}{III}] for each region, which are obtained through the direct method.

Our final compiled sample from the literature contains 174 \ion{H}{II} regions, of which 79 are Galactic and 95 are extragalactic (see Table \ref{tab:literature}). The He$^+$/H$^+$ values for Galactic \ion{H}{II} regions are between $10.18-11.07$, while extragalactic \ion{H}{II} regions values are between $10.70-10.94$ in units of $12+\log_{10}({\rm He}^+/{\rm H}^+)$. The majority of these objects have been used to estimate the primordial helium abundance, as indicated in Sec. \S~\ref{sec:literature}, they are metal-poor \ion{H}{II} regions with O/H between $7.16-8.93$ in units of $12+\log_{10}({\rm O/H})$.

\subsection{Helium abundance He$^+$/H$^+$ in CALIFA \label{sec:He_califa}}

\begin{figure}
\includegraphics[width=\columnwidth]{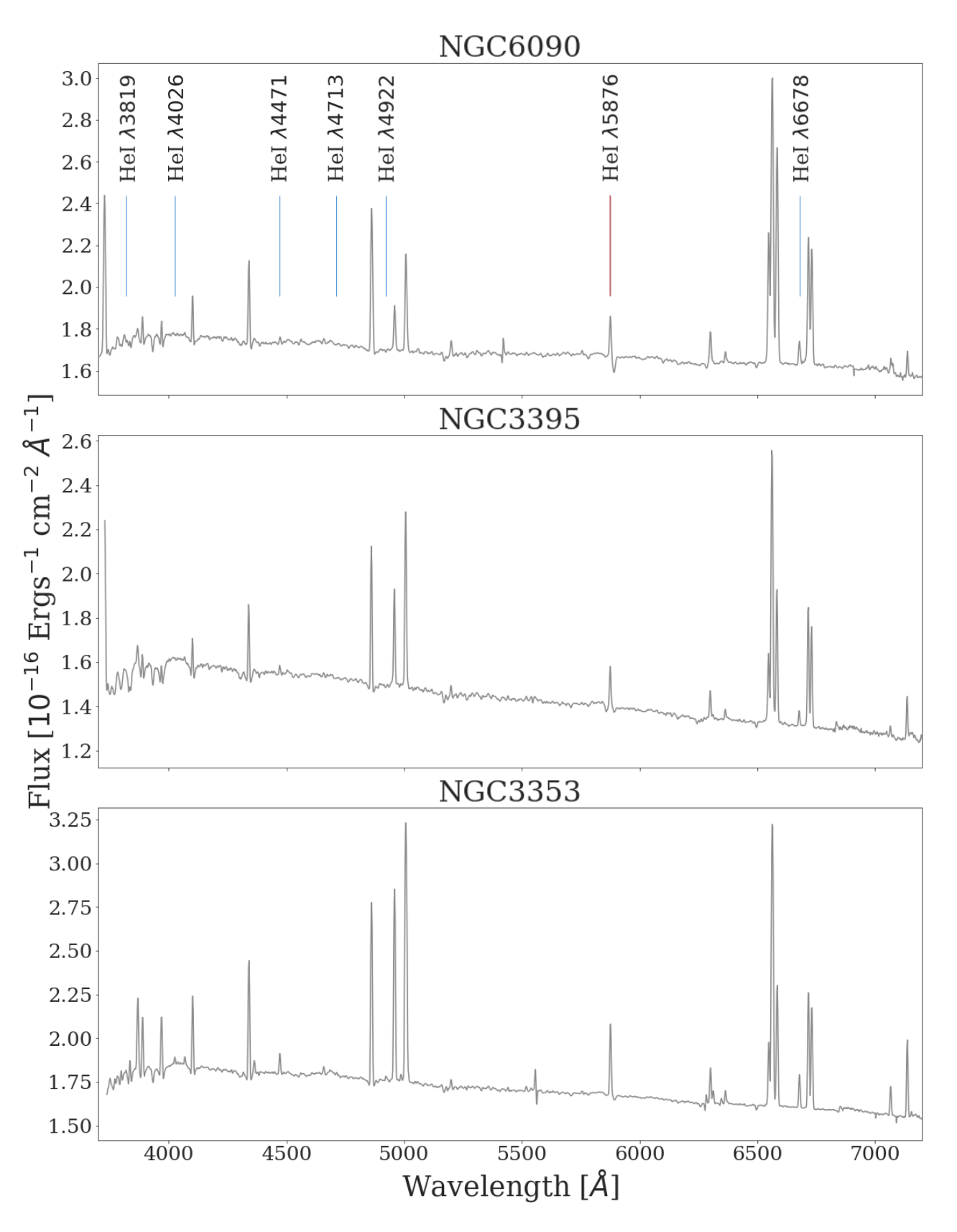}
\caption{Spectra of three \ion{H}{II} regions included in the \citet{Espinosa-Ponce2020} catalog studied along this article, extracted from the CALIFA datacubes, plotted on a logarithmic scale. The identifications of the seven \ion{He}{I} recombination lines contain in the CALIFA catalog: $\lambda3819$, $\lambda4026$, $\lambda4471$, $\lambda4713$, $\lambda4922$, $\lambda5876$, and $\lambda6678$ are indicated at their corresponding wavelength with blue vertical lines, except for $\lambda5876$ red line, which is the most relevant for this work.}
\label{fig:spectra}
\end{figure}

The set of emission lines fluxes included in the \ion{H}{II} regions catalog by \citet{Espinosa-Ponce2020} comprises the following \ion{He}{I} emission lines $\lambda3819$, $\lambda4026$, $\lambda4471$, $\lambda4713$, $\lambda4922$, $\lambda5876$, and $\lambda6678$. Prior to any analysis, we explore the signal-to-noise ratio of these seven \ion{He}{I} emission lines in each cataloged \ion{H}{II} region, and the line with the best S/N is the \ion{He}{I} line $\lambda5876$ (we only use $\lambda5876$ line to develop this work). In Figure \ref{fig:spectra}, we show an example of the spectra corresponding to three \ion{H}{II} regions extracted from the NGC6090, NGC3395, and NGC3553 galaxies. These correspond to those spectra with the best S/N in the \ion{He}{I} line $\lambda5876$ (red vertical line) from the \citet{Espinosa-Ponce2020} catalog. We highlight with vertical lines the seven \ion{He}{I} recombination lines contain in the CALIFA catalog. Our final sample contain those \ion{H}{II} regions where \ion{He}{I} line $\lambda5876$ has S/N$>5$, corresponding to 5386 \ion{H}{II} regions extracted from 465 galaxies.

To calculate the helium abundance, several codes have been designed, including \texttt{PYNEB} \citep{pyneb2015} and \texttt{Helio14} \citep{Peimbert2012}. In a recent work, \citet{Eduardo2020} present the comparison between the helium abundances obtained in both codes for the same set of \ion{He}{I} lines. The general deviation between two methods is less than $0.008\,$dex. In this work, we use \texttt{PYNEB}, an update and expansion code of the \texttt{IRAF} package NEBULAR, which is designed to be more user-friendly and powerful than its predecessors.

We use the version 1.1.13 of \texttt{PYNEB}. The code analyses emission lines from gaseous nebulae and solves the equilibrium equations for an n-level atom. The results depend crucially on the input for the atomic data for each element \citep{pyneb2015}. For recombination lines, \texttt{PYNEB} computes the emissivity of a given line by either interpolating in tables or using a fitting function. We use the effective recombination coefficient computations by \citet{Storey1995} for H lines and \citet{Porter2012,Porter2013} for He lines, which include collisional effects. 

To determine helium abundances, \texttt{PYNEB} uses as an input the electron temperature, electron density, and the intensities of recombination helium lines. Due to the inherent weakness of the emission lines required to estimate the electron temperature, it is not possible to derive this parameter. Therefore, we create a grid in the space defined by these parameters at the specific values $T_e=10,000$, 12,000, 13,000, 15,000, and 20,000 K. correspond to the typical $T_e$ for \ion{H}{II} regions \citep[e.g.][]{Peimbert2012,Izotov2014}. On the other hand, the electron density $n_e$ ($10-250$ cm$^{-3}$) is calculate using the ratio [\ion{S}{II}] $\lambda6717/\lambda6731$. To compute the helium abundance for each \ion{H}{II} regions, we use a combination of temperatures range and adopted as the final value the average of all the estimations. Helium abundance is generally determined by weighting the results from different \ion{He}{I} lines with the brightest line measured has the greatest weight. At visible wavelengths, the brightest \ion{He}{I} recombination line is one of the fine structure \ion{He}{I}, called "Balmer$-\alpha$ of helium". It corresponds to $^3D\rightarrow\,^3P$ ($n=3-2$) transition at 5876\AA\; \citep{Benjamin1999}. For this reason, we decided to use only this line to estimate the final singly ionized helium abundance. In addition, this line has a high probability of being measured in other surveys and therefore can be used to estimate the helium abundance broadly.

We use a MC simulation to estimate the uncertainty of the He$^+$/H$^+$ ratios. We generate 5000 random values for each diagnostic line assuming a Gaussian distribution with a standard deviation equal to the associated uncertainty of the line intensity involved in the diagnostic. In addition, due to the nature of our sample, we incorporate the error due to our inability to constrain $T_e$. As we mentioned before, it is not possible to calculate the electron temperature with the current data. This introduces a systematic uncertainty in our helium abundance estimates due our lack of knowledge of the temperature of each \ion{H}{II} regions. Considering a range between $10,000$ and $20,000$ K in our MC simulation we estimate that this systematic uncertainty is of the order $\sim0.06$ dex. We considered this uncertainty in any further calculation.

\subsection{Helium calibrator \label{sec:calibrator}}

\begin{figure}
\includegraphics[width=\columnwidth]{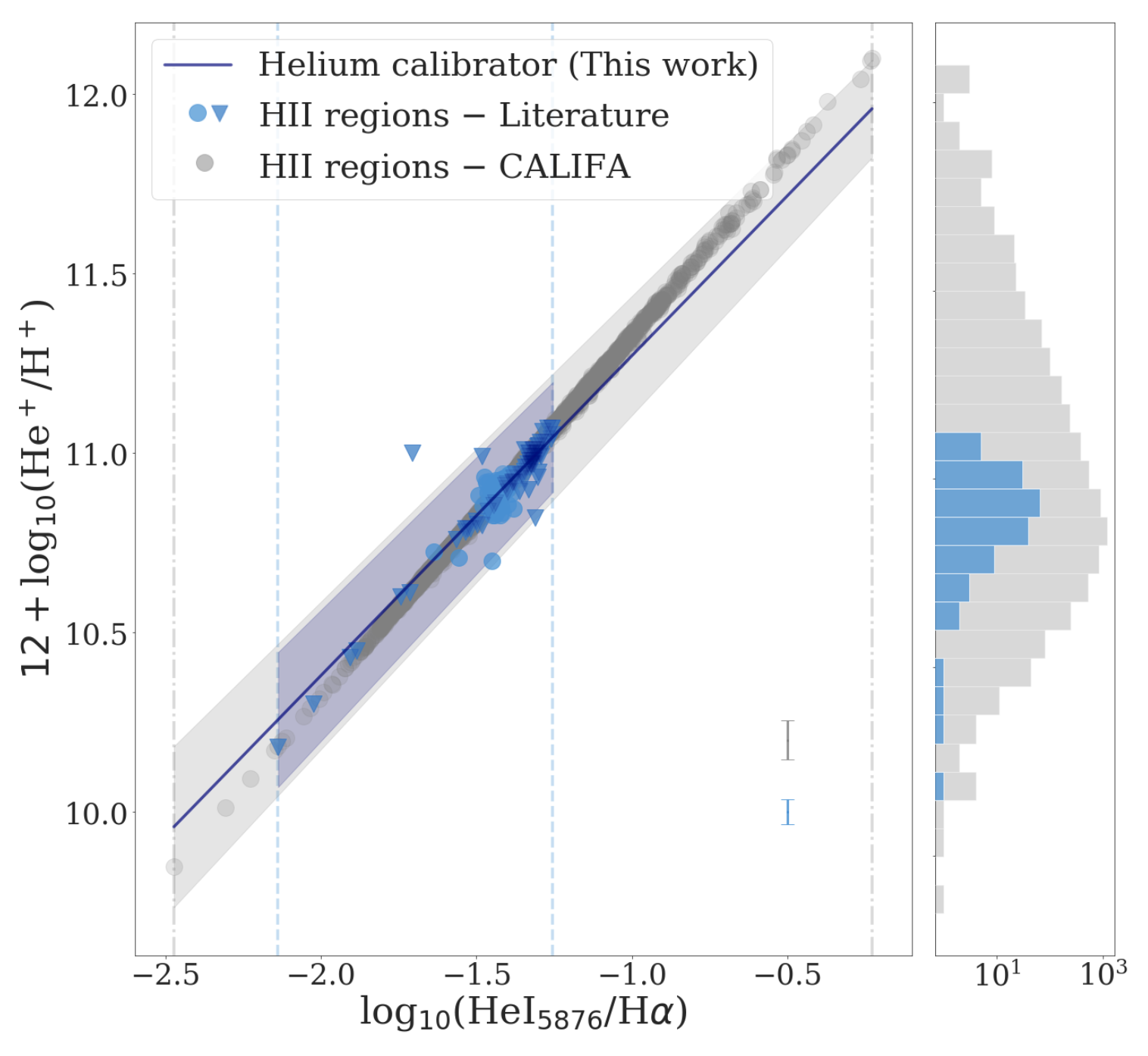}
\caption{Singly ionized helium abundance, $12+\log_{10}({\rm He}^+/{\rm H}^+)$ versus \ion{He}{I} emission line $\lambda5876$ normalized with H$\alpha$, $\log_{10}$(HeI$_{5876}$/H$\alpha)$. In blue color, we show the \ion{H}{II} regions recompiled from literature data, with circles indicating extragalactic and triangles Galactic \ion{H}{II} regions. Gray circles shows singly helium abundances obtained from our CALIFA sample (5386 \ion{H}{II} regions), using only the procedure described in the text (Sec. \ref{sec:He_califa}). The helium calibration is plotted with a dark blue solid line with its applicability interval from $-2.47$ to $-0.23\,$dex. In blue shadow we show the standard deviation for the range between $-2.14$ to $-1.25\,$dex (vertical blue dashed line). In gray shadow we show the standard deviation for the range between $-2.47$ to $-2.14\,$dex and $-1.25$ to $-0.23\,$dex (vertical gray dashed-dotted line). Finally, we show on right lower the average random error associated with the computation of the singly helium abundance for different abundance bins. The vertical histogram represents the distribution of helium abundance for both samples, blue for literature data, and gray for CALIFA data (the number of \ion{H}{II} regions is presented in log scale).}
\label{fig:calibrator}
\end{figure}

We use the sample describe in the Sec. \S~\ref{sec:He_literature} to provide an empirical calibration of the singly ionized helium abundance. In Figure \ref{fig:calibrator} we show helium abundance $12+\log_{10}({\rm He}^+/{\rm H}^+)$ versus the \ion{He}{I} emission line flux $\lambda5876\AA$ for the collection of 95 extragalactic \ion{H}{II} regions (blue circles), and 79 Galactic \ion{H}{II} regions (blue triangles) with singly ionized helium abundance. It is known that the helium abundance and the \ion{He}{I} emission line flux should present a linear dependence when the rest of the physical parameters of the \ion{H}{II} are similar (electron temperature, electron density, the shape of the nebulae). This is indeed the relation predicted by codes like \texttt{PYNEB} \citep{pyneb2015}. Therefore, to obtain our calibrator, we perform a least-squares linear regression to characterize the relation between the two parameters (dark blue line), corresponding to the interval from $-2.14$ to $-1.25\,$dex for $\log_{10}({\rm HeI}_{5876}/{\rm H}\alpha)$ value (vertical blue dashed line). For the current study we perform the largest compilation of public data taking into account the errors introduced by the different observational effects (slit apertures, \ion{H}{II} selection in the different studies, differences in the ionization structures, etc.). All these effects introduce possible systematic offsets between the different datasets. However we do not expect them to strongly bias or affect the final derived relation. On the other hand, we show the singly ionized helium abundance obtained from our sample (5386 \ion{H}{II} regions, gray circles) from CALIFA data, using the procedure describe in the previous section. 

We note that although at a first order a simple linear regression is a good approximation of each of the two analyzed sub-samples (literature and CALIFA \ion{H}{II} regions), this relation does not describe completely well the two samples together. There are some appreciable differences in the observed distribution for the CALIFA and literature data (the difference between both is $\sim0.05\,$dex). In particular, the linear relation estimated based on literature data does not describe the extremes of the dynamical range covered by CALIFA data. On the other hand, the Galactic \ion{H}{II} regions by \citet{Deharveng2000-II}, some points correspond to blue triangles, with $\log_{10}({\rm HeI}_{5876}/{\rm H}\alpha)<-1.7$ do follow the same trend of the CALIFA data (these He abundances are the only ones in the literature that were obtained using just one single \ion{He}{I} line: $\lambda5876$).

The calibrator error is obtained from the difference between the best fit in the literature (calibrator) versus the best fit of the abundances obtained in \ion{H}{II} regions of CALIFA. This way, the final error covers both samples, and this generates a systematic effect as the literature data do not cover the same dynamical range of parameters. The following equation describes the calibrator:

\begin{equation}
12+\log_{10}({\rm He}^+/{\rm H}^+) = x*(0.891 \pm 0.039) + (12.161 \pm 0.055),
\end{equation}

where $x$ is the flux ratio $\log_{10}({\rm HeI}_{5876}/{\rm H}\alpha)$. We report the nominal errors of the coefficients are independent of the range of $x$. However, we have to add a systematic error that is indeed different for different ranges of the $x-$value: i) for the range covered by literature data, (i.e. when $x$ is between $-2.14$ and $-1.25\,$dex), the statistical (total) error is $\pm0.047$ ($\pm0.083$) dex (blue shadow); ii) for values outside this range (from $-2.47$ to $-2.14\,$dex or $-1.25$ to $-0.23\,$dex), the statistical (total) error that must be adopted is $\pm0.053$ ($\pm0.086$) dex (gray shadow). Moreover, we have made a test where we clear outliers from literature (the blue triangles described in the previous paragraph), and we do not observe a significant difference in the fit\footnote{Calibrator without outliers $12+\log_{10}({\rm He}^+/{\rm H}^+) = x*(0.876 \pm 0.056) + (12.139 \pm 0.079)$.}.

It is observed that some of the CALIFA \ion{H}{II} regions present a helium abundance near or even larger than $12.0$ dex. These values are physically not feasible, since it would indicate that helium is more abundant than hydrogen. Due to the systematic and statistical errors, it is possible that the measured values reach that nonphysical regime if their number is low. In Figure \ref{fig:calibrator}, we include a histogram comparing the distributions of the helium abundances for both literature and CALIFA subsamples. As expected, literature data are well peaked at $\sim10.75$ dex, with a tail towards lower values. The CALIFA data present a peak at similar He abundances, but with a wider range of covered values and larger tails. This is a combination of the wider range of \ion{H}{II} region types covered by CALIFA data and the larger systematic and statistical errors. The fraction of regions with high He abundance is very low. Indeed, less than a $0.5\%$ of them have a value larger than $11.75$, and only a $0.03\%$ present unphysical values. We consider that given the described uncertainties this fraction is not abnormal.

\section{Discussion}

In this study, we explore two different datasets, one conformed by a compilation of literature \ion{H}{II} regions with measured He abundances, and another one extracted from the catalog of \ion{H}{II} regions derived from the IFS data provided by the CALIFA data. This is so far the largest extragalactic \ion{H}{II} regions spectroscopic database. Using these data we provide for the first time an empirical calibration for the singly ionized helium abundance describe in Sec. \S~\ref{sec:calibrator}.

To further verify the calibrator, we explore the variations of the helium abundance obtained from our calibrator with respect to the oxygen abundance in both compilations. The oxygen abundances are obtained using different methods; the direct method for the literature data and the strong lines method using the \citet{Ho2019} calibrator for CALIFA data. This strong-line calibrator was derived using an state-of-the-art neural network modelling using several line ratios and anchoring the oxygen abundance to a catalog of \ion{H}{II} regions with values estimated using the direct method.

\begin{figure}
\includegraphics[width=\columnwidth]{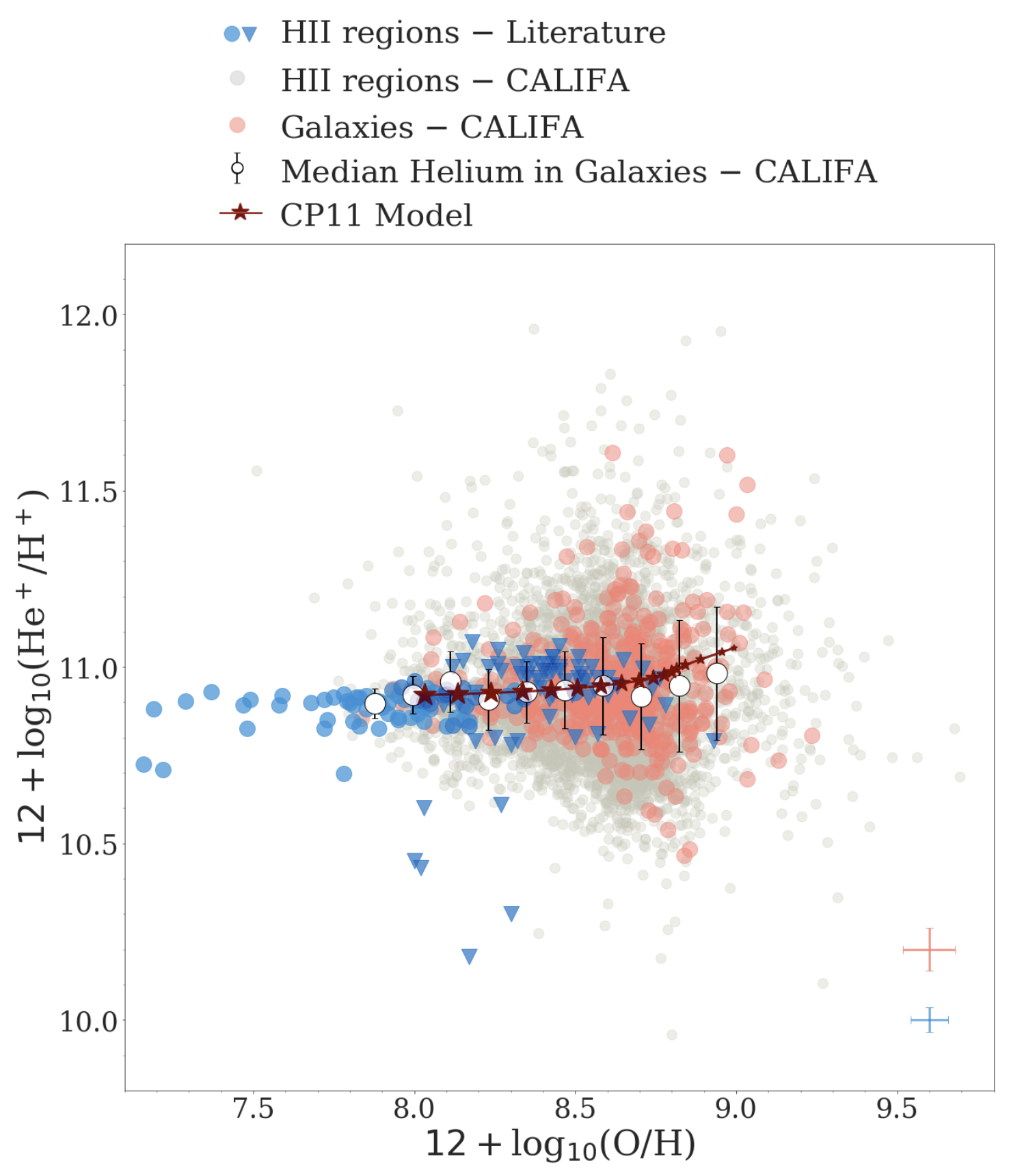}
\caption{Oxygen abundance, $12+\log_{10}({\rm O/H})$ versus singly ionized helium abundance, $12+\log_{10}({\rm He}^+/{\rm H}^+)$. We show (i) the \ion{H}{II} regions recompiled from literature data (Galactic; blue triangles and extragalactic; blue circles); (ii) the individual \ion{H}{II} regions in CALIFA (gray circles); (iii) the average abundances of each galaxy (pink circles); (iv) the median of the helium abundance with a distribution of 10 bins in the oxygen abundance (white circles); (v) the present-time radial profile of the oxygen and helium abundances from 3 to 19 kpc, in 1 kpc steps (red stars) for a chemical evolution model of the Milky Way. The galactocentric distance is in increasing order from right to left as indicated by the size of the stars. On the bottom right, we show the average random error associated with the computation of the abundances.}
\label{fig:OxHe}
\end{figure}

In Figure \ref{fig:OxHe}, we show the oxygen versus helium abundance distribution. For literature data, we show the abundances for each individual Galactic (blue triangles) and extragalactic (blue circles) \ion{H}{II} regions. For the CALIFA data, we present the abundances for each individual \ion{H}{II} regions (gray circles), the average abundances for each galaxy (pink circles), and the average helium abundance from the integrated data within 10 regular bins in oxygen abundance of 0.13 dex width (white circles). We observe that the \ion{H}{II} regions of CALIFA (gray data-points) cover a broader range in oxygen and helium abundances than literature data. CALIFA data covers a metallicity range of $7.51-9.69$ for $12+\log_{10}({\rm O/H})$, while literature \ion{H}{II} regions are located between $7.16-8.93$. As mentioned above (see \S~\ref{sec:literature}), most \ion{H}{II} regions in literature data have been used to determine the primordial helium abundance, and therefore are selected to be metal-poor regions. This may be the reason why there is a bias in helium abundances relative to the literature data in Figure \ref{fig:calibrator} too.

We analyze the median helium abundance in the CALIFA galaxies within 10 bins in oxygen abundance (white circles) in order to obtain the general behavior of helium abundance. Both data from the literature compilation and the \ion{H}{II} regions from CALIFA show a similar trend where at low metallicity ($12+\log_{10}({\rm O/H})<8.5$), the variation of helium abundance is negligible and the scatter around the average in the distribution of helium abundance in galaxies decreases appreciably. And at high metallicity ($12+\log_{10}({\rm O/H})>8.75$), we see that there is an apparent increasing trend for both the median He abundance and the scatter around these values. However, due to the high dispersion in the data, this result, although consistent with what is expected in the models, is not totally conclusive.

To understand the observed trends, we compare them with the values for the oxygen and helium abundances predicted by the inside-out chemical evolution model derived for the Milky Way (MW) described in \citep[hereafter CP11,][]{Carigi2011,Carigi2019}. This model was built to reproduce the radial distributions of the total baryonic mass, and the oxygen abundances of the ISM for different galactocentric radii of our Galaxy. We note that the CP11 model yields a total He/H abundance instead of a He$^+$/H$^+$ one. Nevertheless, it is expected that the difference between the singly ionized helium and total abundances is of the order of $0.008\,$dex, see \citet{Izotov1997,Peimbert2000,Peimbert2012,Valerdi2019,Valerdi2021,Aver2020}. It is known that MW-like galaxies grow from the inside-out following a sequence of SFHs (Star Formation Histories) and ChEHs (CHemical Evolution Histories) at each radii which shape is similar to that galaxies of different masses \citep{Review2020,ReviewII2020}. Thus, a chemical evolution model that reproduces the observed radial gradients in the MW would also recover the differential chemical evolution of galaxies of different masses.

In Figure \ref{fig:OxHe}, we show the present-time oxygen and helium abundances for different galactocentric distances (red stars, from 3 to 19 kpc), derived from the above model and are in increasing order from right to left as indicated by the size of the stars. The CP11 model presents an increasing behavior for helium abundance, mainly for innermost radii corresponding to high metallicity. This is due to the high efficiency in the He production by young, massive, metal-rich stars and old metal-poor low-mass stars. At more recent times, quasi simultaneously, both types of stars enrich the ISM with great amount of helium due to the strong difference in the stellar lifetime of the young-massive stars and the old low mass stars \citep[for more details, see][]{Carigi2008,Romano2010,Carigi2019,K2020}. The observed trend depicts a radial negative gradient for the helium abundance. When compared to the known radial gradient of oxygen abundance \citep[between $-0.04$ and $-0.06\,$dex/kpc][]{Arellano2020} it is clear that the helium gradient is weaker, reaching a mild $-0.01\,$dex/kpc. This trend agrees with the main behavior observed in the average distribution traced by the binned data (white circles). This means, as indicated before, that the MW at different radii behaves like a set of different spiral galaxies of different masses. This is a direct consequence of the connection between the global and resolved trends recently reviewed in \citet{Review2020}.

The increase of He/H with O/H observed for our CALIFA subsample suggests a coupled evolution of both elements. Since the O/H presents a well-known global and local relation with the stellar mass and mass surface density, it is expected that the helium follows a similar relation too. However, as the range of values covered by the Helium abundance is relatively narrower than that of oxygen, it is expected that these relations are shallower than the ones described for O/H.

Helium is not frequently used to trace chemical evolution because the range of evolution is wider in oxygen. However, the joint analysis of the two elements could help to understand the chemical enrichment process in galaxies. CALIFA sample allows us to expand the dynamical range on the exploration of the helium abundance in star-forming regions in galaxies with a larger redshift and wider metallicity ranges. With our sample covering a wide range of galaxy properties, it is possible to constrain chemical evolutionary models at kpc scales.

\section{Conclusions}

We propose an empirical calibration for the singly ionized helium abundance based on a compilation of 174 literature \ion{H}{II} regions. For this purpose we use the most intense \ion{He}{I} emission line at $\lambda5876$. The reported calibrator for this sample is:
\begin{equation}
12+\log_{10}({\rm He}^+/{\rm H}^+) = x*(0.891 \pm 0.039) + (12.161 \pm 0.055),
\nonumber
\end{equation}
where $x$ is the flux ratio $\log_{10}({\rm HeI}_{5876}/{\rm H}\alpha)$ (see Sec \ref{sec:calibrator} for more details). On the other hand, we estimate the singly ionized helium abundance using \texttt{PYNEB} in the largest sample of \ion{H}{II} regions collected from CALIFA survey (5386 \ion{H}{II} regions in 465 galaxies), observed with the technique integral field spectroscopy. Although our calibrator is derived from the literature data, we find that it represents well the two explored samples (literature and CALIFA \ion{H}{II} regions). We use helium abundances estimated from CALIFA data to define the calibrator validity, and estimate its associated error.

We further explore the validity of the calibrator using the distribution of oxygen versus helium abundance (obtained from our calibrator), and we note an apparent increasing behavior for helium abundance for high oxygen abundance. Our result suggests that, at least for the studied sample, the average values for each galaxy are consistent with our understanding of the chemical evolution of the Milky Way.

This is a first attempt of characterization in helium abundance for a large sample of \ion{H}{II} regions located in galaxies representative of the nearby Universe. It would be possible to improve our study with the help of higher resolution instruments, higher quality data, and higher redshift ranges. In this way, we could be obtained a better determination of helium abundance using more than one recombination line.

\section*{Acknowledgements}
We acknowledge support from the grant IA-100420 (DGAPA-PAPIIT, UNAM), PAPIIT-DGAPA-IN100519 project, and funding from the CONACYT grants CF19-39578, CB-285080 and FC-2016-01-1916. M.V. acknowledges to Andr\'es Sixtos for several constructive suggestions in \texttt{PYNEB}. Finally, we would like to acknowledge an anonymous referee for a carefully reading and many helpful suggestions.

\vspace{0.5cm}
{\bf Data Availability.} {We use along this article literature data that are accessible in the quoted articles \citep{Sanchez2012,Sanchez2016a,Sanchez2016b,Sanchez2016DR3}, and data from the \ion{H}{II} regions catalog by \citet{Espinosa-Ponce2020}, publicly available in the following webpage \url{http://ifs.astroscu.unam.mx/CALIFA/HII_regions/}}



\bibliographystyle{mnras}
\bibliography{references} 








\bsp	
\label{lastpage}
\end{document}